\documentclass[conference]{IEEEtran}

\usepackage[T1]{fontenc}
\usepackage{cite}
\usepackage{amsmath,amssymb,amsfonts}
\usepackage{algorithmic}
\usepackage{graphicx}
\usepackage{textcomp}
\usepackage{xcolor}
\usepackage{hyperref}
\usepackage{fancyhdr}
\def\BibTeX{{\rm B\kern-.05em{\sc i\kern-.025em b}\kern-.08em
    T\kern-.1667em\lower.7ex\hbox{E}\kern-.125emX}}

\begin{document}
\newcommand{\todo}[1]{{\textcolor{red}{[#1]}}}

\makeatletter 
\newcommand{\linebreakand}{%
  \end{@IEEEauthorhalign}
  \hfill\mbox{}\par
  \mbox{}\hfill\begin{@IEEEauthorhalign}
}
\makeatother 

\pagenumbering{arabic} 
\pagestyle{fancy}
\fancyfoot[C]{\thepage}

\title{Why Do Experts Favor Solar and Wind as Renewable Energies Despite their Intermittency?}

\author{
\IEEEauthorblockN{Steven P. Reinhardt}
\IEEEauthorblockA{\textit{Transform Computing, Inc.} \\
steve@xfr.ai\\
ORCID 0000-0003-4355-6693}
}

\maketitle

\begin{abstract}

As humanity's shift to renewable energy generation accelerates, people who are
not experts in renewable energy are learning about energy technologies and the energy market, which are complex.
The answers to some questions will be obvious to expert practitioners but not
to non-experts.
One such question is Why solar and wind generation are expected
to supply the bulk of future energy when they are  intermittent.
We learn here that once the baseline hurdles of scalability to utility scale and
the underlying resources being widely available globally are satisfied, 
the forecasted cost of solar and wind is 2-4X lower than competing
technologies, even those that are not as scalable and available.
The market views intermittency  as surmountable.

\end{abstract}


\section{Introduction}
\label{sec:introduction}

As a relative newcomer to the field of renewable energy and microgrids, I have many questions, some high-level and some detailed.  
Recently I was trying to understand why there is so much focus on solar and wind as high-value renewable energies despite their intermittency, and I could not find a good answer via web search or chatbot prompt.  
A few hours searching revealed the answer to that question and a few others that came up during the search, which this note documents.

Humankind  faces the urgent need to avoid the Earth's climate breakdown, which would be catastrophic for human society.
Reducing greenhouse-gas emissions from energy production is a major portion of that task, so worldwide there is a massive shift from greenhouse-gas-emitting fuels to \emph{renewable} sources of energy; examples include solar, wind, geothermal, hydro, biomass, ocean, hydrogen, concentrated solar, and nuclear, with specific technologies spanning a wide range of maturity.  
At Transform Computing (XFR), we develop explainable AI (XAI) for microgrids and so we need to understand  microgrid configurations and capabilities deeply.  Which sources of renewable energy will be most commonly chosen for microgrids?  What are the strengths and weaknesses of those renewable-energy sources that can be exploited and mitigated?  How can a weakness of one technology be offset by a strength of another?  Which technologies will be deployed soon at utility scale?

The answer is that two compelling factors  favor  solar and wind: a) they are widely and abundantly available and b) they   generate energy at least as cheaply now as any  alternative and are expected to be even cheaper in the future.  The magnitude of these advantages makes energy planners  accept  solar's and wind's intermittency and the cost and complexity required to store the generated energy during abundant generation times until heavy consumption times.

\begin{figure*}[!tbp]
  \centering
  \begin{minipage}[b]{0.34\textwidth}
    \includegraphics[width=\textwidth]{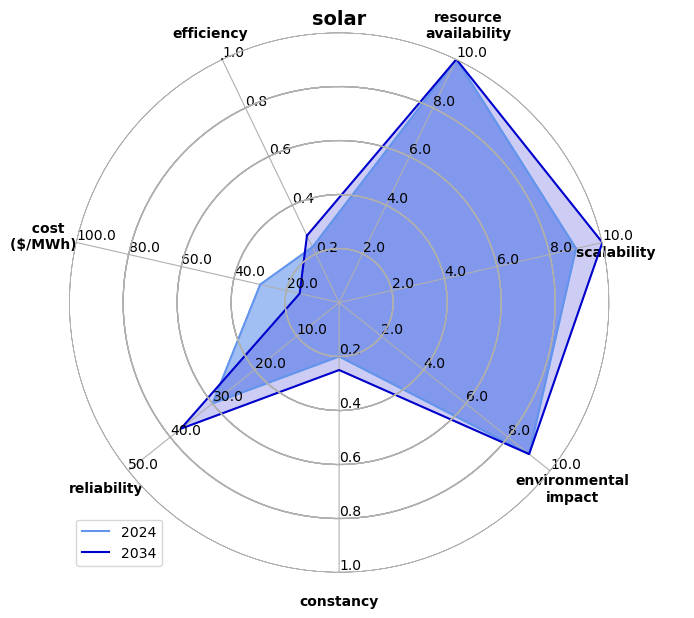}
  \end{minipage}
  \begin{minipage}[b]{0.34\textwidth}
    \includegraphics[width=\textwidth]{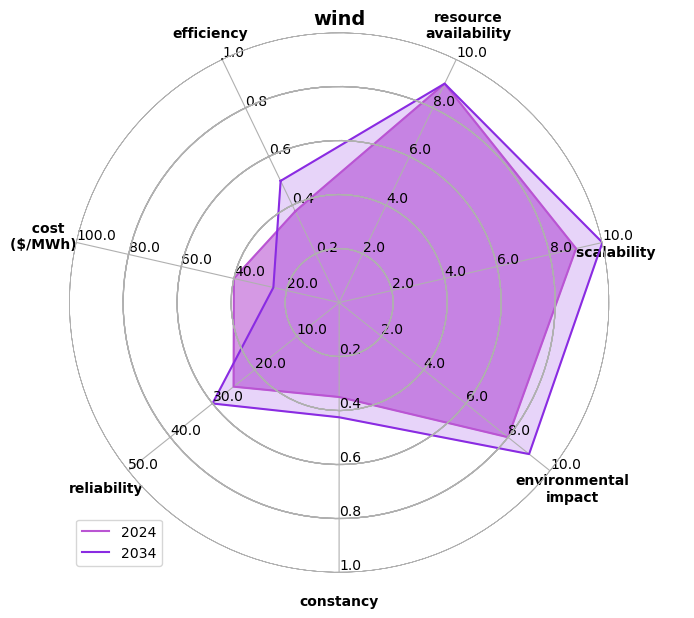}
  \end{minipage}
  \begin{minipage}[b]{0.34\textwidth}
    \includegraphics[width=\textwidth]{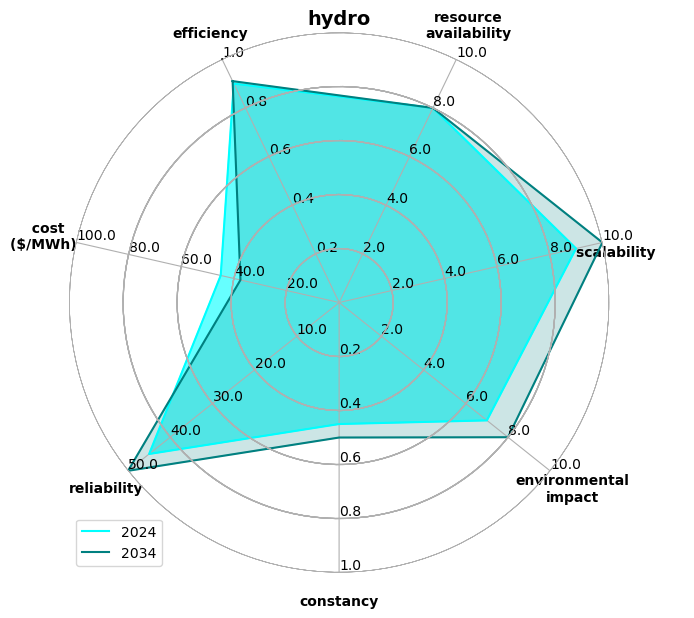}
  \end{minipage}
    \begin{minipage}[b]{0.34\textwidth}
    \includegraphics[width=\textwidth]{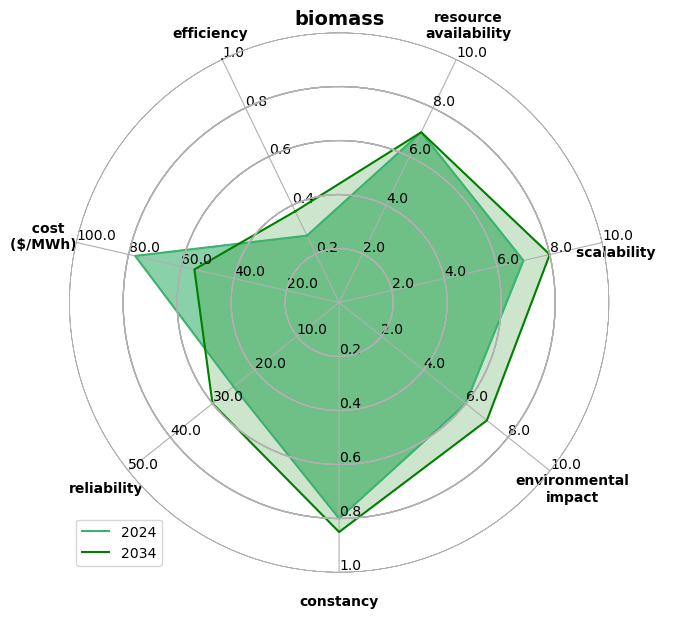}
  \end{minipage}
  \begin{minipage}[b]{0.34\textwidth}
    \includegraphics[width=\textwidth]{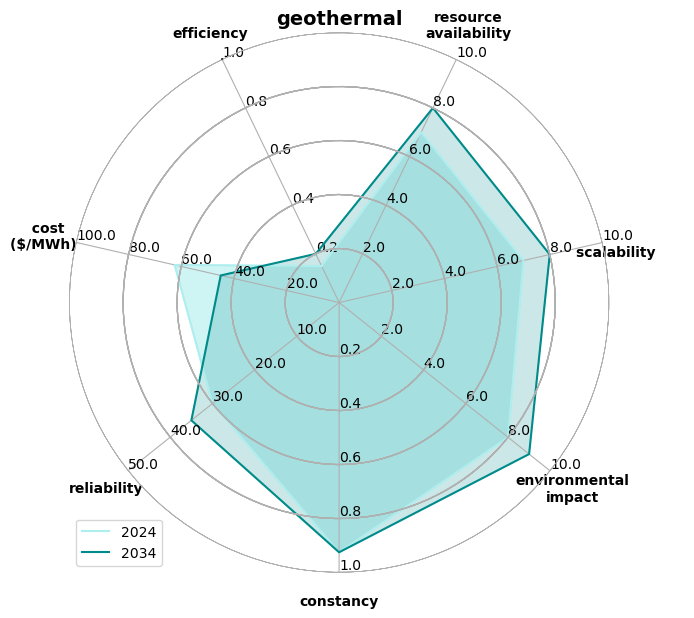}
  \end{minipage}
  \begin{minipage}[b]{0.34\textwidth}
    \includegraphics[width=\textwidth]{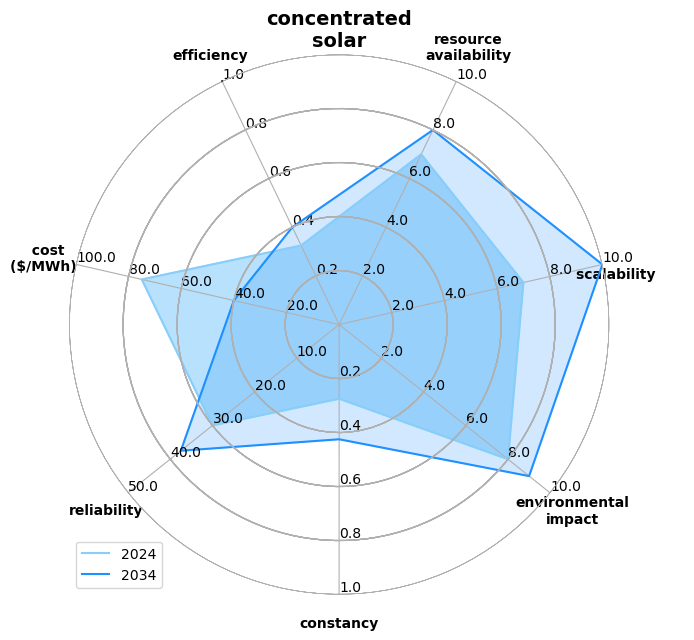}
  \end{minipage}
  \begin{minipage}[b]{0.34\textwidth}
    \includegraphics[width=\textwidth]{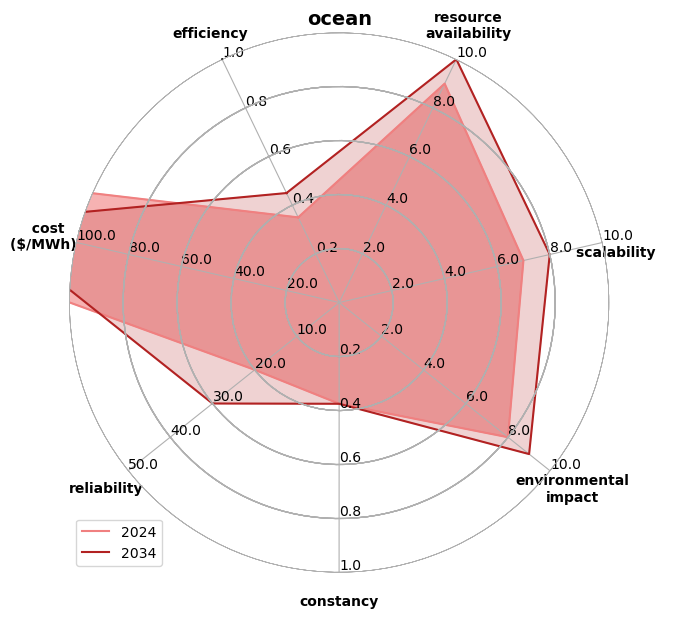}
  \end{minipage}
    \begin{minipage}[b]{0.34\textwidth}
    \includegraphics[width=\textwidth]{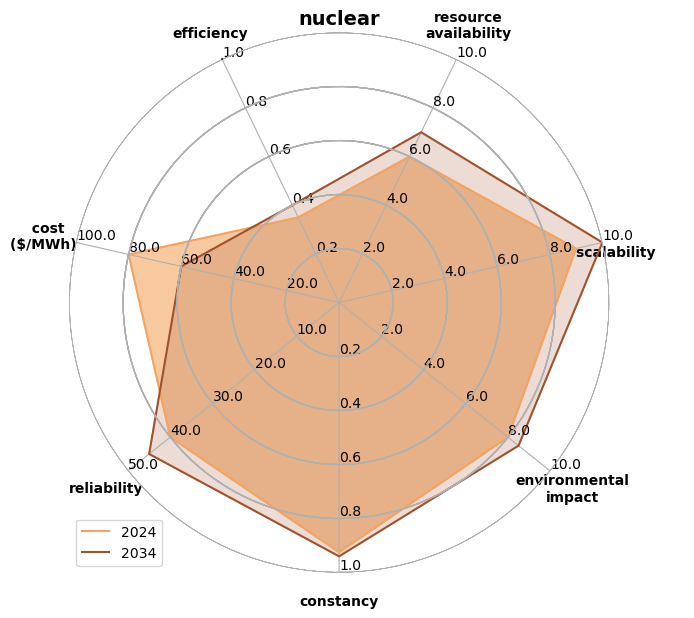}
  \end{minipage}
  \caption{Current (2024) and 10-year-future (2034) predicted performance of renewable-energy-generation technologies.}
  \label{fig:2024_2034_radar_plots}
  \end{figure*}
  
\section{Key Dimensions of  Energy Sources}
\label{sec:important_dimensions}
Searching the literature for analyses of renewable technologies quickly surfaces many technology dimensions that determine their value for utility-scale energy generation.  
We prioritize by the most important dimensions of technologies that will be in the market in a few years, namely these seven:
\begin{itemize}
  \item \textbf{Efficiency}: The raw efficiency with which a technology converts the natural phenomenon to energy is important in the early stages of a technology because it is used by potential funders to make funding decisions.  It can also be important in later stages, as a technology with widespread adoption that still has low efficiency and hence considerable upside may be attractive.  Otherwise, raw efficiency is not important from a practical perspective.  Efficiency is measured as a ratio between 0.0 and 1.0, where 1.0 is perfect efficiency.
  \item \textbf{Cost}: For the definition of cost, we use 
  \emph{levelized cost of energy} 
  \cite{doe_lcoe}
  which includes both initial costs, including financing, and recurring costs; the units are US dollars per megawatt hour (\$/MWh); smaller is better.
  \item \textbf{Reliability}: Focusing on utility-scale deployment, reliability covers a variety of issues, from  device physics to wear and tear in a physical environment, such as exposure to sun or salt water.  We include stability as an aspect of reliability; if a technology varies through time or from device to device in a way that cannot be readily mitigated, that reduces its reliability.  Reliability is measured as mean time to interruption (MTTI) in years.
  \item \textbf{Constancy}: Constancy is the inverse of intermittency.  What fraction of a day or week will the technology generate energy?  E.g., solar is determined by when the sun is shining and wind  by when the wind is blowing.  
  Constancy is measured on a scale of 0.0 to 1.0.
  \item \textbf{Environmental Impact}:  This includes both impact during the fabrication (e.g., use of rare-earth metals by solar panels) and construction (e.g., destruction of habitat by hydro) phases as well as during operation.  Almost no technology has zero environmental impact; this  is about trade-offs.  Environmental impact is measured on a scale of 1 (worst) to 10 (best).
  \item \textbf{Scalability}: We define scalability as whether the technology can be practically expanded from the lab or prototype stage to utility scale.  Again we use a scale of 1 to 10. 
  \item \textbf{Resource Availability}: This means whether the needed natural resource is sufficiently available.  For example, the most exploitable  geothermal reservoirs, those with \emph{high} temperature, often defined as above 150\textdegree C, exist in less than 10 percent of Earth's land area 
  \cite{ucs_geothermal}, 
  so resource availability is not ideal.  We use a scale of 1 to 10.
\end{itemize}

\section{Current and 10-year-future Performance}
\label{sec:current_10yearfuture_perf}
I chose to compare solar, wind, hydro, biomass, geothermal, concentrated solar\footnote{Concentrated solar captures solar energy as heat, so is more \emph{dispatchable} than photovoltaic (PV) solar, but is currently less mature.}, ocean, and nuclear energies\footnote{Of course, nuclear energy is not \emph{renewable} in the normal meaning of that term,
but it is a proven utility-scale  energy source that does not emit greenhouse gases and so it is often being considered as an option.}.
I chose to measure each technology on the seven dimensions both now (summer 2024), 
for which we have fairly good data, and 10 years in the future (2034).
Of course, predicted trajectories that far into the future
will likely be wrong in detail, but can still offer high-level guidance;
governments, utilities,  other companies, and individuals are 
making large-scale investment decisions now based on their view of these trajectories,
and multiple organizations are predicting future costs for renewable generation
(see Section \ref{ssec:discussion_closer_cost}), so it does not seem premature to collect 
those predictions in one place.

\subsection{Reading the radar charts}
The \emph{radar} or Kiviat plots are in Figure \ref{fig:2024_2034_radar_plots}.  
\begin{itemize}
    \item All  dimensions are "bigger is better" except cost.
    \item In each plot, the less transparent polygon represents 2024 and the more transparent polygon represents 2034.
    \item The cost axis is  limited to \$100/MWh,
    so the lowest costs (below \$50/MWh) are distinguishable.
    Ocean costs of \$225/MWh (2024) and \$115 (2034) are not visible. 
\end{itemize}

\begin{figure*}[!tbp]
  \centering
  \begin{minipage}[b]{1.0\textwidth}
    \includegraphics[width=\textwidth]{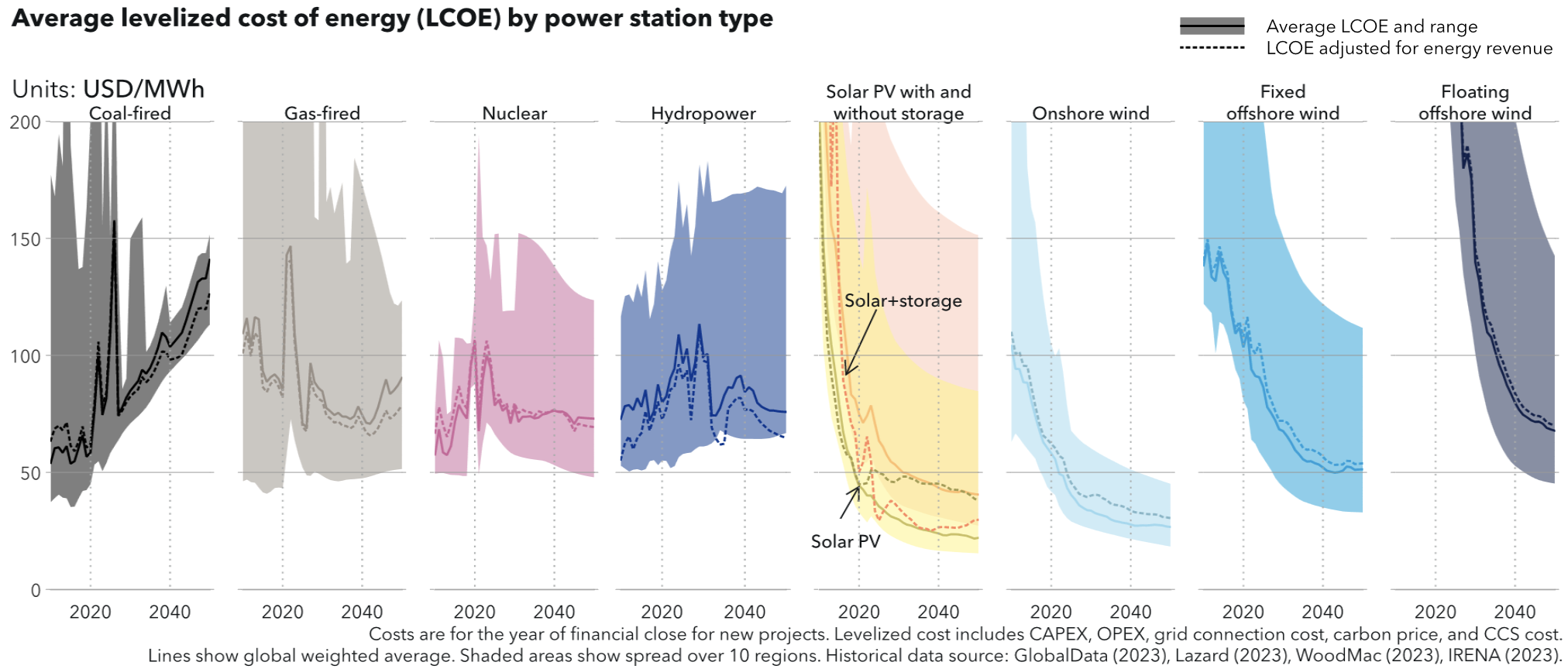}
  \end{minipage}
 
  \caption{Renewable-energy levelized cost of energy (LCOE) predictions from  
  \cite{dnv_energy_transition}
  }
  \label{fig:DNV_Energy_transition_outlook_costs}
  \end{figure*}

\section{Per-Technology Analysis}
\label{sec:per-techno_analysis}

\subsection{Solar}
Solar has the best cost of any of these technologies, along 
with superb environmental impact, scalability, and resource availability.  
It is intermittent.  
It is also not that absolutely efficient, which means it has 
potential upside.
\subsection{Wind}
We do not split out on-shore wind from off-shore wind; on-shore is more mature
but faces more opposition from landowners.  Wind has excellent cost, 
and environmental impact, scalability, and resource availability nearly as good
as solar's, along with solar's intermittency.  
It is also not that absolutely efficient, which could be an opportunity.
Its reliability is not as strong as the best technologies.
\subsection{Hydro}
Hydro is a mature technology with strong efficiency, good cost, and strong reliability, 
so I wondered why it is not discussed more commonly as a major source of future renewable generation.
Its plot reveals two key issues:  it has good cost now, but that cost is not expected to decrease much,
and its resource availability, esp. in the US, is limited (i.e., many of the best locations
are already  built).
\subsection{Biomass}
Biomass is an emerging technology that has not been widely deployed at utility scale, 
partly due to cost, efficiency,  reliability, resource availability, 
and environmental impact deficits compared to the best technologies.  
Its efficiency is still low and can be viewed as an opportunity.
\subsection{Geothermal}
Residential heat pumps are a geothermal technology already commonly installed in areas that do not have the coldest temperatures.  Higher temperature is widely viewed as essential to expanding the use of heat pumps, even beyond residential, and new technologies (e.g., super-deep geothermal) could expand the limited range noted above.
\subsection{Ocean}
Ocean technologies harness the motion of water, notably tidal.  
This is another emerging technology, whose cost has not yet made it practical
but which is receiving substantial government investment.  
If lower-cost implementations  enable scalability, the ubiquity
of oceans would make it attractive.
\subsection{Nuclear} 
Despite nuclear energy being a proven technology, albeit with a few  severe failures,
its cost remains unattractively high, 
and its resource availability does not compare with 
the best technologies.
The US Department of Energy (DoE) is funding development of advanced nuclear reactor technologies that could overcome current cost and manufacturability deficits. 

\section{Discussion}
\label{sec:discussion}

\subsection{The technologies are markedly different}
One of the things that the radar format makes  obvious is how
different the technologies are. 
Solar and wind are cheap, abundant, and intermittent; note they are
relatively mature within this context.  Cost varies by a factor of 5-6 between cheapest (solar) and costliest (ocean).  
Biomass, geothermal, concentrated solar, and ocean are still nascent 
and need to prove
themselves.

\subsection{Once scalability and availability are high, cost is critical}
Comparing geothermal to solar and wind may be  illuminating.  
Geothermal is highly constant, but  
its availability is only good, 
and its cost is 3-5X higher 
than solar and wind, which limits its role, given the current technology trajectory.
Note also that solar and wind  recently achieved lower cost than natural gas, 
often viewed as the cheapest greenhouse-gas-emitting source, and that by 2034
they are forecasted to have a 2X cost advantage.  
If these forecasts match future reality,  
the status quo  of continuing with natural gas will not be
economically viable.
Wind and photovoltaic solar benefit from manufacturing economies of scale, 
which so far nuclear and hydro do not; new nuclear and hydro technologies
with higher manufacturability could reduce their costs.
\subsection{The market believes that intermittency can be overcome}
As this paper's title says, the big negative for solar and wind is their intermittency.
Governments and companies  are still depending on them to
play a big role in renewable generation.  It appears the market believes
that  intermittency can be overcome, presumably by  storage 
to shift the energy to  when it is needed.
That said, today's energy storage does not have the duration (days or weeks) or cost (\$20/kWh 
\cite{ziegler2019storage},
compared to 2024's  \$150/kWh) we need for grid scaling, so further development is vital.
\subsection{Some level of constant generation is valuable}
Detailed analyses have shown that while the intermittency of solar and wind
is not a major issue now, it will make shifting the last 20\% or 10\%
of generation to renewable sources unacceptably expensive 
\cite{sepulveda2018role}
Thus we can expect a role for constant technologies even if more costly.
\subsection{Looking more closely into cost}
\label{ssec:discussion_closer_cost}
Figure \ref{fig:DNV_Energy_transition_outlook_costs} shows a detailed cost forecast for many
of our candidate technologies, on the same \$/MWh scale, further into the future (2050).  
Importantly, it shows progression over time
and a range of values, not the simplistic single value (for just 2024 and 2034) shown in Figure \ref{fig:2024_2034_radar_plots}.
We  observe that the mature technologies (coal, gas, nuclear, and hydro) are forecasted
to have only small cost reductions (and coal's LCOE  increases markedly), 
where the less mature technologies are forecasted to continue to reduce in cost significantly.
Note that the solar subplot includes not only solar by itself (grey lines) 
but also combined solar+storage (red lines), addressing the reality that the non-constancy of solar must
be paired with storage to make it useful in the real world.  
Even with storage costs, solar is roughly tied with onshore wind for most cost-effective technology
(wind is also non-constant and  needs storage).

Another useful source for cost predictions is DoE's National Renewable
Energy Lab's (NREL) Annual Technology Baseline 
\cite{nrel_annual_tech_baseline}
that can be parameterized  to forecast future power costs for renewable and non-renewable technologies.

\subsection{Looking more closely into environmental impact}

Of all the dimensions used here, environmental impact  may have the fuzziest measurement,
as there are many potential environmental impacts, some newly realized, and some whose magnitudes
are hard to assess.  
Two useful surveys that mention many factors but do not try to weight them or coalesce them into a single rating
as in this work are by 
Guidi et al. \cite{guidi2023environmental}
, notably Table 16, and 
Osman et al. \cite{osman2023cost} see Table 5.
Our ratings are necessarily somewhat subjective and more useful to compare technologies relatively than absolutely.

\subsection{Combining constant and intermittent renewable energies}
Another question I had is under what circumstances a developer of a generation plant primarily using a constant renewable technology (e.g., geothermal, biomass) would choose to incorporate (intermittent) solar or wind 
generation as well.
They do not seem a good way to supply power during 
maintenance or unexpected failures, as using an intermittent 
technology for that purpose seems to require the use of batteries anyway, 
as a backup to the backup.
Given solar and wind's cost advantage, it may be that the question 
is best asked the other way around -- 
under what circumstances to include constant technologies? -- 
with one answer being specific use cases where constancy is paramount.

\subsection{Discovery continues}
With the importance of avoiding climate breakdown, many scientists 
worldwide are looking for ways to eliminate greenhouse-gas emissions and 
find and transport energy by different media, so we can expect that
new ideas will continue to emerge.
A recent example of this is 
geologic hydrogen \cite{hand2023hidden},
which is created by natural
mechanisms and resides subsurface in volumes that could contribute greatly 
to our energy supply.

\section*{Conclusions} 
Solar and wind generation are scalable to utility scale 
and the underlying resources are globally widely available.
Despite their being intermittent, 
they are valuable because their cost, even including storage 
to overcome the intermittency, is superior (by 2X or more) to the alternatives, 
including both greenhouse-gas-emitting and renewable technologies.
Digging deeper, cost-effective constant renewable technologies 
will be valuable in eliminating the last 10\% of greenhouse-gas-emitting
generation, and much cheaper and longer-duration storage is needed for grid scale.

\section*{Acknowledgments} 
Benjamin Weiner's comments greatly improved this paper.

\bibliographystyle{IEEEtran}  
\bibliography{references}  



\end{document}